\documentclass[seceq]{ptptex}
\newcommand{\D}{\displaystyle}

\usepackage{epsfig}
\usepackage{graphicx}

\title{
Relativistic Mean Field Theory for Deformed Nuclei with Pairing
Correlations}
\author{
Lisheng Geng$^{1,3,}$\footnote{E-mail: lsgeng0@rcnp.osaka-u.ac.jp}
Hiroshi Toki$^{1,2,}$\footnote{E-mail: toki@rcnp.osaka-u.ac.jp}
\\ Satoru Sugimoto$^{2,}$\footnote{E-mail: satoru@riken.go.jp} and Jie Meng$^{3,}$ \footnote{E-mail: mengj@pku.edu.cn} }
\inst{
$^1$Research Center for Nuclear Physics (RCNP), Osaka University,\\
Ibaraki 567-0047, Japan
\\
$^2$The Institute of Physical and Chemical Research (RIKEN),\\
Wako 351-0198, Japan\\
$^3$School of Physics, Peking University, Beijing 100871, P. R.
China }
\abst{
We develop a relativistic mean field (RMF) description of deformed
nuclei with pairing correlations in the BCS approximation. The
treatment of the pairing correlations for nuclei whose Fermi
surfaces are close to the threshold of unbound states needs
special attention.  With this in mind, we use a delta function
interaction for the pairing interaction to pick up those states
whose wave functions are concentrated in the nuclear region and
employ the standard BCS approximation for the single-particle
states obtained from the RMF theory with deformation.  We apply
the RMF + BCS method to the Zr isotopes and obtain a good
description of the binding energies and the nuclear radii of
nuclei from the proton drip line to the neutron drip line. }
\begin{document}
\maketitle
\section{Introduction}
It is our strong desire to obtain a model valid for all nuclei,
including unstable ones, from the proton drip line to the neutron
drip line. The relativistic mean field (RMF) theory has been used
to describe such nuclei with one parameter set in all mass regions
\cite{hirata92,sugahara94,ring}.  We have to include deformation
and pairing correlations into the RMF model for a suitable
description of finite nuclei.  There is an extended study of all
the even-even nuclei over the entire mass region by Hirata et al.
including only deformation \cite{hirata97}. This calculation
provides a good account of all the nuclei and indicates that
almost all nuclei, except for those with most of the magic
numbers, are deformed. Their calculation, however, does not
include pairing correlations, because the conventional BCS
treatment with the constant pairing interaction is not able to
treat the case in which the Fermi surface is close to the unbound
threshold \cite{hirata97}.

The pairing correlations, on the other hand, have been treated
nicely in the framework of the relativistic Hartree-Bogoliubov
(RHB) method by Meng et al. \cite{JP.96,meng98}.  They were able
to treat nuclei near the unbound threshold in the RHB framework.
They solved the RHB equation in the coordinate space with box
boundary conditions. This method has been applied to many proton
magic nuclei and is capable of describing many interesting
features, such as the giant halo where the neutron density
distribution extends far from the nuclear region. This method is,
however, limited to spherical nuclei. When extended to deformed
system, this method requires such great compuatation time that we
are not able to make calculations for all the nuclei in the
periodic table in a systematic manner.

Recently, there was an interesting suggestion made by Yadav et al.
that using the delta function interaction in the BCS formalism
with proper box boundary conditions can provide a good description
of the proton magic nuclei \cite{yadav02}.  Indeed, the calculated
results obtained from the RMF+BCS method and those obtained in the
RHB framework are nearly equal for those proton magic nuclei with
spherical shapes. The justification of this method was provided by
Sandulescu et al. by solving the resonance states and taking into
account the width effect exactly in the non-relativistic Skyrme
Hartree-Fock framework\cite{sand.00}. The applicability and
justification of such a delta function interaction is also
discussed extensively in a work by Dobaczewski et
al.\cite{Dobac.96} and references therein, where it is shown in
the context of HFB calculations that the use of a delta force in
finite space simulates the effect of finite range interaction in a
phenomenological manner and can take into account the effects of
unbound states properly. It is very interesting, therefore, to
apply this prescription to deformed nuclei.

The self-consistent RMF theory was first extended to treat
deformed nuclei by Price et al.\cite{price.87} and Gambhir et
al.\cite{Gambhir90}. In Gambhir's work, the nucleon wave functions
and the meson fields are expanded in terms of the harmonic
oscillator wave functions. We employ this method for the mean
field part and replace the BCS part with a constant pairing
interaction by one with a delta function interaction. The fact
that we use the nuclear wave functions (in particular, we compute
the overlap between occupied and unoccupied states),  allows us to
pick up states in the continuum whose wave functions are
concentrated in the nuclear region.  This method, at least, is
effective for spherical nuclei and may also be effective for the
deformed case. We mention that the relativistic Hartree-Bogoliubov
theory was worked out using the expansion method for deformed
nuclei \cite{GAL.99}.

In this paper, we formulate the RMF theory with the BCS method for
pairing correlations.  In \S2, we present the RMF formalism with
deformation and pairing.  In \S3, we apply the method to Zr
isotopes from the proton drip line to the neutron drip line. We
compare the calculated results with those obtained with the
assumption of spherical shapes in RCHB and those obtained without
including the pairing correlations. In \S4, we provide the summary
of the present work.

\section{RMF with deformation and pairing}

We present here the formulation of the RMF theory with deformation
and pairing correlations.  We employ the model Lagrangian density
with nonlinear terms for both $\sigma$ and $\omega$ mesons, as
described in detail in Ref. \cite{sugahara94}, which is given by
\begin{equation}
\begin{array}{lll}
{\cal L} &=& \bar \psi (i\gamma^\mu\partial_\mu -M) \psi +
\,\frac{\D 1}{\D 2}\partial_\mu\sigma\partial^\mu\sigma-\frac{\D
1}{\D 2}m_{\sigma}^{2} \sigma^{2}- \frac{\D 1}{ \D
3}g_{2}\sigma^{3}-\frac{\D 1}{\D
4}g_{3}\sigma^{4}-g_{\sigma}\bar\psi
\sigma \psi\\
&&-\frac{\D 1}{\D 4}\Omega_{\mu\nu}\Omega^{\mu\nu}+\frac{\D 1}{\D
2}m_\omega^2\omega_\mu\omega^\mu +\frac{\D 1}{\D
4}g_4(\omega_\mu\omega^\mu)^2-g_{\omega}\bar\psi
\gamma^\mu \psi\omega_\mu\\
 && -\frac{\D 1}{\D 4}{R^a}_{\mu\nu}{R^a}^{\mu\nu} +
 \frac{\D 1}{\D 2}m_{\rho}^{2}
 \rho^a_{\mu}\rho^{a\mu}
     -g_{\rho}\bar\psi\gamma_\mu\tau^a \psi\rho^{\mu a} \\
      && -\frac{\D 1}{\D 4}F_{\mu\nu}F^{\mu\nu} -e \bar\psi
      \gamma_\mu\frac{\D 1-\tau_3}{\D 2}A^\mu
      \psi,\\
\end{array}
\end{equation}
where the field tensors of the vector mesons and of the
electromagnetic field take the following forms:
\begin{equation}
\left\{
\begin {array}{rl}
\Omega_{\mu\nu} =&
\partial_{\mu}\omega_{\nu}-\partial_{\nu}\omega_{\mu},\\
 R^a_{\mu\nu} =& \partial_{\mu}
                  \rho^a_{\nu}
                  -\partial_{\nu}
                  \rho^a_{\mu}-2g_\rho\epsilon^{abc}\rho^b_\mu\rho^c_\nu,\\
 F_{\mu\nu} =& \partial_{\mu}A_{\nu}-\partial_{\nu}
A_{\mu},
\end{array}\right.
\end{equation}
and other symbols have their usual meanings.

The classical variational principle leads to the Dirac equation,
\begin{equation}\label{eq:dirac}
\left[ -i\alpha \nabla + V({\bf r}) + \beta \left(M+S({\bf
r})\right) \right] ~\psi_{i} = ~\epsilon_{i} \psi_{i},
\end{equation}
for the nucleon spinors and the Klein-Gordon equation,
\begin{equation}\label{eq:gordon}
\left\{\begin{array}{lll} \{ -\Delta + m_{\sigma}^{2}
\}\sigma({\bf r})
 &=& -g_{\sigma}\rho_{s}({\bf r})
-g_{2}\sigma^{2}({\bf r})-g_{3}\sigma^{3}({\bf r}),\\

\  \{ -\Delta + m_{\omega}^{2} \} \omega^{\mu}({\bf r})
&=& g_{\omega}j^\mu({\bf r}) + g_4 \omega^2_\mu ({\bf r})\omega^\mu({\bf r}), \\

\  \{ -\Delta + m_{\rho}^{2} \}\rho^{a\mu}({\bf r})
&=& g_{\rho} j^{a\mu}({\bf r}),\\

\  -\Delta A^\mu({\bf r}) &=& ej^\mu_p({\bf r}),
\end{array}\right.
\end{equation}
for the mesons. Here, $V({\bf r})$ represents the vector potential
\begin{equation}
V({\bf r}) = g_{\omega} \gamma^\mu\omega_\mu({\bf r}) +
g_{\rho}\tau^{a}\gamma^\mu\rho^a_\mu({\bf r}) + e\frac{\D
1-\tau_{3}}{ \D 2} \gamma^\mu A_\mu({\bf r}),
\end{equation}
and $S({\bf r})$ is the scalar potential
\begin{equation}
S({\bf r}) = g_{\sigma} \sigma({\bf r}).
\end{equation}

For the mean field, the nucleon spinors provide the corresponding
source terms:
\begin{equation}
\left\{
\begin{array}{lll}
\rho_{s}({\bf r}) &=& \sum\limits_{i=1}^{A} \bar\psi_{i}~\psi_{i},\\

j^\mu({\bf r}) &=& \sum\limits_{i=1}^{A} \bar\psi_{i}\gamma^\mu\psi_{i},\\

j^{a\mu}({\bf r}) &=&\sum\limits_{i=1}^{A}
\bar\psi_{i}\gamma^\mu\tau^a\psi_{i},
\\

j^\mu_p({\bf r}) &=& \sum\limits_{i=1}^{A}
\bar\psi_i\gamma^\mu\frac{\D 1-\tau_{3}}{\D 2}\psi_{i}.
\end{array}\right.
\end{equation}
Here, the summations are taken over the valence nucleons only. It
should be noted that as usual, the present approach ignores the
contribution of negative energy states (i.e. no-sea
approximation), which implies that the vacuum is not polarized.
The coupled equations (\ref{eq:dirac}) and (\ref{eq:gordon}) are
non-linear quantum field equations, and their exact solutions are
very complicated. For this reason, the mean field approximation is
generally used; i.e., the meson field operators in Eq.
(\ref{eq:dirac}) are replaced by their expectation values. In this
treatment, the nucleons are considered to move independently in
the classical meson fields. The coupled equations are solved
self-consistently by iteration.

The symmetries of the system simplify the calculations
considerably. In all the systems considered in this work, there
exists time reversal symmetry, so there are no currents in the
nucleus and therefore the spatial vector components of
$\omega^\mu$, $\rho^{a\mu}$ and $A^\mu$ vanish. This leaves only
the time-like components, $\omega^0$, $\rho^{a0}$ and $ A^0$.
Charge conservation guarantees that only the 3-component of the
isovector $\rho^{00}$ survives.

\subsection{Axially symmetric case}

The RMF theory was extended to treat deformed nuclei with axially
symmetric shapes by Gambhir et al. \cite{Gambhir90}. To make clear
the notation used, here a brief review of the RMF method for
axially deformed nuclei is given.

Many deformed nuclei can be described with axially symmetric
shapes. In this case, rotational symmetry is lost, and therefore
the total angular momentum, $j$, is no longer a good quantum
number. However, the densities are still invariant with respect to
rotation about the symmetry axis, which is assumed to be the
$z$-axis in the following. It is then useful to work with
cylindrical coordinates: $x=r_\bot\cos \varphi,
y=r_\bot\sin\varphi$ and $z$. For such nuclei, the Dirac equation
can be reduced to a coupled set of partial differential equations
in the two variables $z$ and $r_\bot$. In particular, the spinor
$\psi_i$ with the index $i$ is now characterized by the quantum
numbers $\Omega_i, \pi_i$ and $t_i$, where
$\Omega_i=m_{l_i}+m_{s_i}$ is the eigenvalue of the symmetry
operator $J_z$, $\pi_i$ is the parity and $t_i$ is the isospin.
The spinor can be written in the form
\begin{equation}\label{eq:spinor}
\psi_i({\bf r},t)=\left(\begin{array}{c}f_i({\bf r})\\ig_i({\bf
r})\end{array}\right) =\frac{\D
1}{\D\sqrt{2\pi}}\left(\begin{array}{l}f^+_i(z,r_\bot)e^{i(\Omega_i-1/2)\varphi}\\
f^-_i(z,r_\bot)e^{i(\Omega_i+1/2)\varphi}\\ig^+_i(z,r_\bot)e^{i(\Omega_i-1/2)\varphi}\\
ig^-_i(z,r_\bot)e^{i(\Omega_i+1/2)\varphi}\end{array}\right)\chi_{t_i}(t).
\end{equation}
The four components $f^\pm_i(z,r_\bot)$ and $g^\pm_i(z,r_\bot)$
obey the coupled Dirac equations. For each solution with positive
$\Omega_i$, $\psi_i$, we have the time-reversed solution with the
same energy, $\psi_{\bar{i}}=T\psi_i$, with the time reversal
operator $T=-i\sigma_y K$ ($K$ being the complex conjugation). For
nuclei with time reversal symmetry, the contributions to the
densities of the two time reversed states, $i$ and $\bar{i}$, are
identical. Therefore, we find the densities
\begin{equation}
\rho_{s,v}=2\sum_{i>0}\left((|f^+_i|^2+|f^-_i|^2)\mp(|g^+_i|^2+|g^-_i|^2)\right)
\end{equation}
and, in a similar way, $\rho_3$ and $\rho_c$. The sum here runs
over only states with positive $\Omega_i$. These densities serve
as sources for the fields $\phi=\sigma, \omega^0, \rho^{00}$ and
$A^0$, which are determined by the Klein-Gordon equation in
cylindrical coordinates.

To solve the RMF equations, the basis expansion method is used. We
closely follow the details, presentation and notation of Ref.
\cite{Ring.97}. For the axially symmetric case, the spinors
$f_i^\pm$ and $g_i^\pm$ in Eq. (\ref{eq:spinor}) are expanded in
terms of the eigenfunctions of a deformed axially symmetric
oscillator potential,
\begin{equation}
V_{\textrm{osc}}(z,r_\bot)=\frac{1}{2}M\omega_z^2
z^2+\frac{1}{2}M\omega^2_{\bot}r^2_\bot.
\end{equation}
Then, imposing volume conservation, the two oscillator frequencies
$\omega_\bot$ and $\omega_z$ can be expressed in terms of a
deformation parameter, $\beta_0$: $\omega_z=\omega_0
\exp\left(-\sqrt{\frac{5}{4\pi}}\beta_0\right)$ and
$\omega_\bot=\omega_0
\exp\left(+\frac{1}{2}\sqrt{\frac{5}{4\pi}}\beta_0\right)$.

The basis is now determined by the two constants $\omega_0$ and
$\beta_0$.  The eigenfunctions of the deformed harmonic oscillator
potential are characterized by the quantum numbers,
$|\alpha\rangle=|n_z,n_r,m_l,m_s\rangle$, where $m_l$ and $m_s$
are the components of the orbital angular momentum and of the spin
along the symmetry axis. The eigenvalue of $J_z$, which is a
conserved quantity in these calculations, is $\Omega=m_l+m_s$. The
parity is given by $\pi=(-)^{n_z+m_l}$.

The eigenfunctions of the deformed harmonic oscillator can be
written explicitly as
\begin{equation}
\Phi_\alpha(z,r_\bot,\varphi,s,t)=\phi_{n_z}(z)\phi_{n_r}^{m_l}
(r_\bot)\frac{\D1}{\D\sqrt{2\pi}}e^{im_l\varphi}\chi_{m_s}(s)\chi_{t_\alpha}(t),
\end{equation}
with
\begin{equation}
\begin{array}{l}
\phi_{n_z}(z)=\frac{\D
N_{n_z}}{\D\sqrt{b_z}}H_{n_z}(\zeta)e^{-\zeta^2/2},\\
\phi_{n_r}^{m_l}(r_\bot)=\sqrt{2}\frac{\D N_{n_r}^{m_l}}{\D
b_\bot}\eta^{m_l/2}L_{n_r}^{m_l}(\eta)e^{-\eta/2},
\end{array}
\end{equation}
where $\zeta=z/b_z$ and $\eta=r^2_\bot/b^2_\bot$. The polynomials
$H_n(\zeta)$ and $L_n^m(\eta)$ are the Hermite polynomials and the
associated Laguerre polynomials, as defined in Ref. \cite{AS.70}.
The quantities $N_{n_z}$ and $N_{n_r}^{m_l}$ are normalization
constants.

The spinors $f_i^\pm$ and $g_i^\pm$ in Eq. (\ref{eq:spinor}) are
explicitly given by the following relations:
\begin{equation}
\left\{
\begin{array}{lll}
f^+_i(z,r_\bot)&=&\sum\limits_\alpha^{\alpha_{\textrm{max}}}f_\alpha^{(i)}\phi_{n_z}(z)\phi_{n_r}^{(\Omega-1/2)}(r_\bot),\\
f^-_i(z,r_\bot)&=&\sum\limits_\alpha^{\alpha_{\textrm{max}}}f_\alpha^{(i)}\phi_{n_z}(z)\phi_{n_r}^{(\Omega+1/2)}(r_\bot),\\
g^+_i(z,r_\bot)&=&\sum\limits_\beta^{\beta_{\textrm{max}}}g_\beta^{(i)}\phi_{n_z}(z)\phi_{n_r}^{(\Omega-1/2)}(r_\bot),\\
g^-_i(z,r_\bot)&=&\sum\limits_\beta^{\beta_{\textrm{max}}}g_\beta^{(i)}\phi_{n_z}(z)\phi_{n_r}^{(\Omega+1/2)}(r_\bot).
\end{array}
\right.
\end{equation}
The quantum numbers $\alpha_{\textrm{max}}$ and
$\beta_{\textrm{max}}$ are chosen in such a way that the
corresponding major quantum numbers $N=n_z+2n_\rho+m_l$ are not
larger than $N_F+1$ for the expansion of the small components and
not larger than $N_F$ for the expansion of the large components.

\subsection{Pairing with delta function interaction}

Based on the single-particle spectrum calculated with the RMF
method described  above, we carry out a state-dependent BCS
calculation \cite{Lane.64,Ring.80}. The gap equation has a
standard form for all the single particle states,
\begin{equation}\label{eq:bcs}
\Delta_k=-\frac{\D 1}{\D 2}\sum_{k'>0}\frac{\D \bar{V}_{kk'}
\Delta_{k'}}{\D\sqrt{(\varepsilon_{k'}-\lambda)^2+\Delta_{k'}^2}},
\end{equation}
where $\varepsilon_{k'}$ is the single-particle energy and
$\lambda$ is the Fermi energy. The particle number condition is
given by $2\sum\limits_{k>0} v_k^2=N$. In the present work, we use
a delta force for the pairing interaction,
\begin{equation}
V=-V_0\delta({\bf r_1}-{\bf r_2}),
\end{equation}
with the same strength $V_0$ for both protons and neutrons. The
pairing matrix element for the $\delta$-function force is given by
\begin{equation}\label{eq:me}
\begin{array} {lll}\bar{V}_{ij}&=&\langle i\bar{i}|V|j\bar{j}\rangle-\langle i\bar{i}|V|\bar{j}j\rangle=-V_0\int d^3r\,
\left[\psi^\dagger_i\psi^\dagger_{\bar{i}}\psi_j\psi_{\bar{j}}-\psi^\dagger_i\psi^\dagger_{\bar{i}}\psi_{\bar{j}}\psi_j\right]\\
\end{array},
\end{equation}
with the pairing energy defined by
$E_{\textrm{pair}}=-\sum\limits_{k>0}\Delta_k u_k v_k$.  Equations
(\ref{eq:dirac}) and (\ref{eq:gordon}), the gap equations
(\ref{eq:bcs}), and the total particle number condition $N$ for a
given nucleus are solved self-consistently by iteration.

\section{Numerical calculation for Zr isotopes}

We apply the formalism to the Zr isotopes from the proton drip
line to the neutron drip line.  For the RMF Lagrangian, we use the
$A$-dependent parameter set TMA
\cite{sugahara94,THSST.95,sugahara95}. The parameter values are as
follows. The masses of nucleon, $\sigma$, $\omega$ and $\rho$
mesons are, respectively, $M=938.900 \mbox{ MeV}$,
$m_\sigma=519.151 \mbox{ MeV}$, $m_\omega=781.950 \mbox{ MeV}$,
$m_\rho=768.100 \mbox{ MeV}$. The effective strengths of the
couplings between various mesons and nucleons have the values
$g_\sigma=10.055+3.050/A^{0.4}$, $g_\omega=12.842+3.191/A^{0.4}$
and $g_\rho=3.800+4.644/A^{0.4}$. The nonlinear coupling strengths
of the $\sigma$ meson are given by
$g_2=-0.328-27.879/A^{0.4}\mbox{ (}\mbox{fm}^{-1}\mbox{)}$, and
$g_3=38.862-184.191/A^{0.4}$, whereas the self-coupling of the
$\omega$ field has the strength $g_4=151.590-378.004/A^{0.4}$. For
the pairing interaction, we take the strength of the delta
function interaction as $V_0=343.7 \mbox{ MeV fm}^{3}$, which was
obtained by requiring that the experimental value of the proton
pairing gap in $^{90}$Zr (1.714 MeV) be reproduced with a given
energy cutoff ($E_{\textrm{max}}-\lambda\le 8.0$ MeV).

Because in many nuclei we have several solutions at different
equilibrium deformations with similar energies, it is difficult to
select the ground-state configuration uniquely. The procedure we
employed is as follows. The basis deformation $\beta_0$ is set
equal to $\beta_{2m}$, following the results of Hirata et al.
\cite{hirata97}, in which a constrained calculation
\cite{Flocard.73} was carried out for the quadrupole moment,
$Q_{20}$, to obtain the lowest minimum in the energy curve of each
nucleus.

The present calculation was performed by expansion in 14
oscillator shells for both the fermion fields and the boson
fields. The convergence of this calculation has been tested with
20 shells for both the fermion fields and the boson fields.
Following Ref. \cite{Gambhir90}, we fix $\hbar\omega_0=41A^{-1/3}$
for fermions. In what follows, we discuss the details of our
calculations and the numerical results for the Zr isotopes.
\begin{table}[htbp]
\caption{The ground state properties of even Zr isotopes
calculated with the parameter set TMA. Listed are the total
binding energy, $B_{tot}$, the binding energy per nucleon,
$B_{per}$ , charge, neutron, proton, and matter root mean square
radii, $R_c$, $R_n$, $R_p$ and $R_m$, and the quadrupole
deformation parameter for the neutron, proton and matter
distributions, $\beta_{2n}$, $\beta_{2p}$ and $\beta_{2m}$, with
$A$ the mass number and $N$ the neutron number.}
\begin{center}\label{table1}
\begin{tabular}{cc@{\hspace{2ex}}|@{\hspace{2ex}}cc@{\hspace{2ex}}|@{\hspace{2ex}}cccc@{\hspace{2ex}}|@{\hspace{2ex}}ccc}
\hline\hline
$A$&$N$&$B_{tot}$&$B_{per}$&$R_c$&$R_n$&$R_p$&$R_m$&$\beta_{2n}$&$\beta_{2p}$&$\beta_{2m}$\\
 \hline\hline
 78&38&639.089&8.193&4.3331&4.1496&4.2586&4.2058&0.480&0.507&0.494\\
 80&40&667.668&8.346&4.3431&4.2048&4.2688&4.2369&0.489&0.506&0.498\\
 82&42&692.014&8.439&4.4085&4.3134&4.3353&4.3241&0.589&0.579&0.584\\
 84&44&716.742&8.533&4.2734&4.2081&4.1979&4.2033&-0.205&-0.210&-0.207\\
 86&46&739.502&8.599&4.2694&4.2381&4.1938&4.2175&-0.148&-0.166&-0.156\\
 88&48&762.784&8.668&4.2628&4.2618&4.1871&4.2280&0.002&0.003&0.002\\
 90&50&784.859&8.721&4.2697&4.2969&4.1941&4.2515&0.000&0.000&0.000\\
 92&52&797.819&8.672&4.3056&4.3678&4.2307&4.3087&-0.121&-0.137&-0.128\\
 94&54&811.838&8.637&4.3504&4.4372&4.2762&4.3695&0.216&0.226&0.220\\
 96&56&825.419&8.598&4.3832&4.5023&4.3095&4.4230&0.267&0.269&0.268\\
 98&58&837.427&8.545&4.5171&4.6629&4.4456&4.5755&0.525&0.517&0.522\\
 100&60&849.997&8.500&4.4694&4.6457&4.3972&4.5480&0.412&0.393&0.405\\
 102&62&860.800&8.439&4.4896&4.6945&4.4177&4.5880&0.407&0.391&0.401\\
 104&64&870.744&8.373&4.5117&4.7424&4.4402&4.6285&0.406&0.395&0.402\\
 106&66&879.994&8.302&4.5357&4.7903&4.4646&4.6701&0.409&0.403&0.407\\
 108&68&888.649&8.228&4.5579&4.8366&4.4872&4.7102&0.408&0.407&0.408\\
 110&70&895.861&8.144&4.5930&4.8991&4.5228&4.7657&0.442&0.435&0.440\\
 112&72&902.253&8.056&4.6302&4.9699&4.5606&4.8277&0.501&0.466&0.488\\
 114&74&910.199&7.984&4.5182&4.9054&4.4468&4.7496&-0.169&-0.173&-0.170\\
 116&76&916.310&7.899&4.5284&4.9396&4.4571&4.7788&-0.142&-0.156&-0.147\\
 118&78&922.451&7.817&4.5189&4.9614&4.4475&4.7934&0.001&0.000&0.000\\
 120&80&928.345&7.736&4.5364&4.9908&4.4653&4.8220&-0.001&-0.001&-0.001\\
 122&82&933.816&7.654&4.5542&5.0184&4.4834&4.8495&0.000&0.000&0.000\\
 124&84&933.842&7.531&4.5653&5.0909&4.4947&4.9065&-0.014&-0.009&-0.013\\
 126&86&934.295&7.415&4.5980&5.1619&4.5279&4.9694&0.162&0.115&0.147\\
 128&88&935.359&7.307&4.6488&5.2217&4.5794&5.0298&0.237&0.204&0.227\\
 130&90&936.150&7.201&4.6746&5.2808&4.6057&5.0826&0.261&0.228&0.251\\
 132&92&937.086&7.099&4.6606&5.3443&4.5914&5.1278&-0.232&-0.184&-0.218\\
 134&94&937.686&6.998&4.6787&5.3932&4.6098&5.1718&-0.239&-0.188&-0.224\\
 136&96&937.995&6.897&4.6960&5.4382&4.6274&5.2128&-0.241&-0.190&-0.226\\
 138&98&938.148&6.798&4.7133&5.4808&4.6449&5.2522&-0.243&-0.190&-0.227\\
 140&100&937.988&6.700&4.7314&5.5211&4.6632&5.2902&-0.244&-0.192&-0.229\\
 \hline\hline
\end{tabular}
\end{center}
\end{table}

\subsection{Binding energy per nucleon and two neutron separation energy}
\begin{figure}\epsfxsize = 10 cm \centerline{
\includegraphics[width=0.8\textwidth]{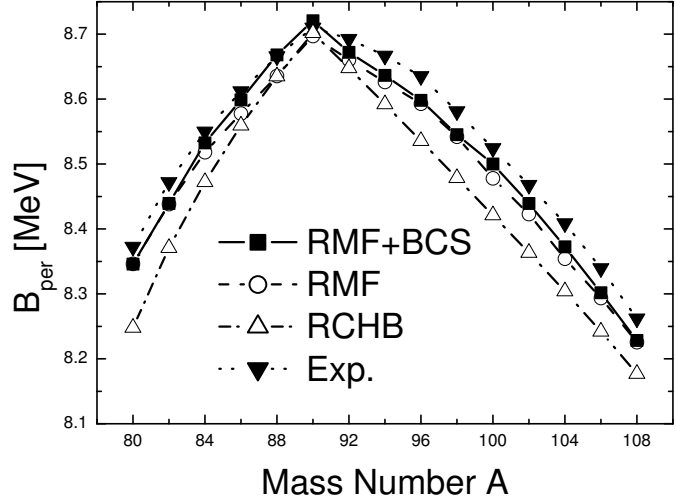}}
\caption{\label{fig1.fig} Binding energy per nucleon, $B_{per}$,
for even Zr isotopes as functions of mass number $A$ obtained from
the deformed RMF+BCS calculations (squares), the deformed RMF
calculations (circles), the spherical RCHB calculations
\cite{MR.98} (up triangles) and the experimental data
\cite{Audi.95} (down triangles).}
\end{figure}
\begin{figure}\epsfxsize = 10 cm\centerline{
\includegraphics[width=0.8\textwidth]{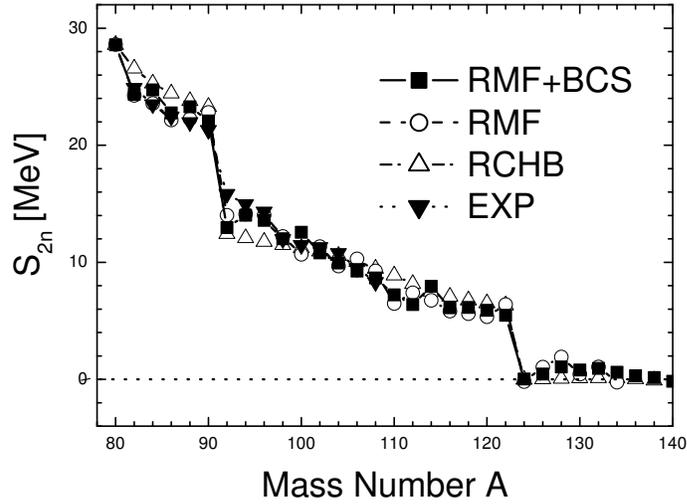}}
\caption{\label{fig2.fig} Two neutron separation energies,
$S_{2n}$, for even Zr isotopes as functions of mass number $A$
obtained from the deformed RMF+BCS calculations (squares), the
deformed RMF calculations (circles), the spherical RCHB
calculations \cite{MR.98} (up triangles) and the experimental data
\cite{Audi.95} (down triangles).}
\end{figure}
The two neutron separation energy,  $S_{2n}$, defined as
\begin{equation}
S_{2n}(Z,N)=B(Z,N)-B(Z,N-2),
\end{equation}
is quite a sensitive quantity to test a microscopic theory, where
$B(Z,N)$ is the binding energy of nuclei with proton number Z and
neutron number N.  The two neutron separation energy becomes
negative when the nucleus becomes unstable with respect to
two-neutron emission. Hence, the drip line nucleus for the
corresponding isotope chain is the one with two less neutrons than
the nucleus at which $S_{2n}$ first becomes negative.

In Figs. \ref{fig1.fig} and \ref{fig2.fig},  we plot the results
for the binding energy per nucleon of $^{80-108}$Zr and the
results for the two-neutron separation energies for the entire
chain of Zr isotopes covering the proton and neutron drip lines.
The figures also display the results of the RMF calculations, the
results of the spherical RCHB calculations \cite{MR.98} and the
available experimental data\cite{Audi.95}. First, from Fig.
\ref{fig1.fig}, we can see that the RMF+BCS calculations give a
better description of the binding energy per nucleon than the RMF
calculations. The largest difference between the RMF+BCS results
and the experimental values is less than $0.04$ MeV. Noting here
that most of the nuclei are deformed and that despite this fact,
we did not readjust any parameter for our calculations, the
agreement is quite remarkable. Second, from Fig. \ref{fig2.fig},
we see basically good agreement between experiment and the present
calculation. The values of $S_{2n}$ for the so-called giant halos
\cite{MR.98} ($^{124}$Zr$-$$^{138}$Zr) are reproduced accurately.
The strong variation in the experimental separation energy at the
neutron magic number $N=50$ is well accounted for by the present
calculation. The small staggering for $^{84}$Zr, $^{88}$Zr,
$^{100}$Zr and $^{114}$Zr can be attributed to a mixture of the
pairing interaction and the deformation effect. With the
assumption of spherical shapes, it is predicted in Ref.
\cite{MR.98} that the drip-line nucleus for Zr isotopes is
$^{140}$Zr. We also obtain $^{140}$Zr as the drip-line nucleus. In
both Figs. \ref{fig1.fig} and \ref{fig2.fig}, the deformation
effect is clearly seen. The deformed calculations describe the
experimental data much better than the spherical calculations.
This once again points out the need for an appropriate
relativistic calculation with both deformation and proper pairing
interaction taken into account in order to obtain a reliable
description of all the nuclei from the proton drip line to the
neutron drip line.

\subsection{Root mean square neutron radii}

The root mean square neutron radius is another basic important
physical quantity to describe neutron-rich nuclei. In the mean
field theory, the root mean square (rms) neutron radii can be
directly deduced from the neutron density distributions, $\rho_n$:
\begin{equation}
R_n=\langle r^2_n\rangle^{1/2}= \left[\frac{\int \rho_n r^2d{\bf
r}}{\int \rho_n d{\bf r}}\right]^{1/2}
\end{equation}

\begin{figure}
\epsfxsize = 10 cm \centerline{
\includegraphics[width=0.8\textwidth]{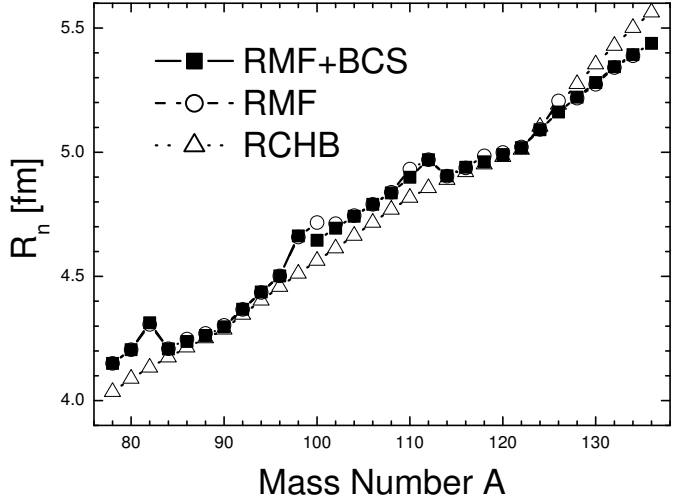}}
\caption{\label{fig3.fig} The root mean square neutron radii for
even Zr isotopes as functions of mass number $A$ obtained from the
deformed RMF+BCS calculations (squares), the deformed RMF
calculations (circles), and the spherical RCHB calculations
\cite{MR.98} (triangles).}
\end{figure}

In Fig. \ref{fig3.fig}, the root mean square neutron radii for Zr
nuclei are presented. Although the calculation was done with the
assumption of spherical shapes, the results of RCHB \cite{MR.98}
are also shown for comparison. Two interesting features are
clearly seen. First, the so-called giant halos
($^{124}$Zr$-$$^{138}$Zr)\cite{MR.98} are obtained.  Second,
nuclei with large absolute values $\beta_{2m}$, $^{78}$Zr$-$$
^{82}$Zr and $^{94}$Zr$-$$^{112}$Zr (see also Table. \ref{table1}
and Fig. \ref{fig6.fig}), tend to have larger rms neutron radii.
This can also be seen from the fact that some RMF results (for
$^{100}$Zr and $^{126}$Zr) are larger than the RMF+BCS results.

Another point to note here is that despite the second effect
mentioned above, for the so-called giant halos, the present work
gives relatively small values for the root mean square neutron
radii. We believe that this is due to the harmonic oscillator
basis we used in the calculations of the deformed nuclei. This
deficiency will be resolved in the near future, although a great
deal of effort is necessary to change the numerical method.
Nevertheless, despite the small discrepancies for giant halos, the
present work provides good agreement with the RCHB\cite{MR.98}
results and can give a reliable prediction of the rms neutron
radii for all the nuclei from the proton drip line to the neutron
drip line.

\subsection{Single-particle states and their occupation
probabilities}

One interesting feature of exotic nuclei is the contribution from
the continuum due to the pairing correlations. In this context, it
is very interesting to study the amount by which the contributions
from the continuum differ for the calculations with a constant
pairing interaction and the calculations with a delta function
interaction.
\begin{figure}
\epsfxsize = 10 cm \centerline{
\includegraphics[width=0.8\textwidth]{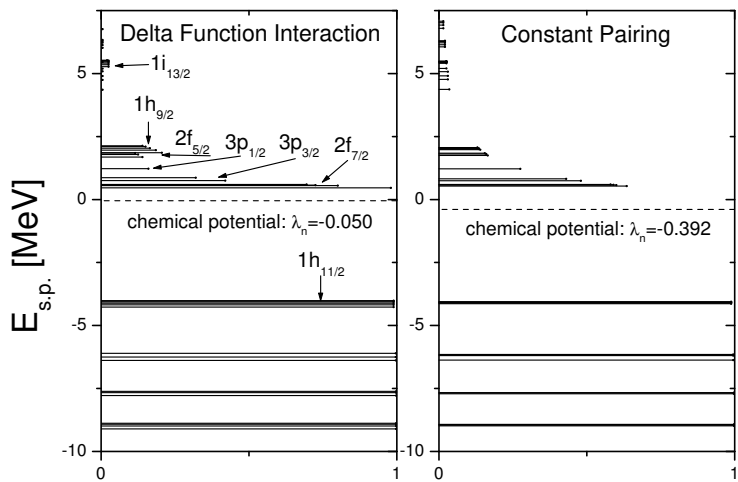}}
\caption{\label{fig4.fig} The occupation probability is
represented by solid horizontal bar for each neutron single
particle state of $^{124}$Zr. The left panel displays the results
of the RMF+BCS calculation with the delta function interaction,
while the right panel displays the results of the constant pairing
calculation. The dotted horizontal line represents the Fermi
energy. In both panels, the occupation probabilities of the
continuum states are multiplied by a factor of 5 for clarity.}
\end{figure}
\begin{figure}
\epsfxsize = 10 cm \centerline{
\includegraphics[width=0.8\textwidth]{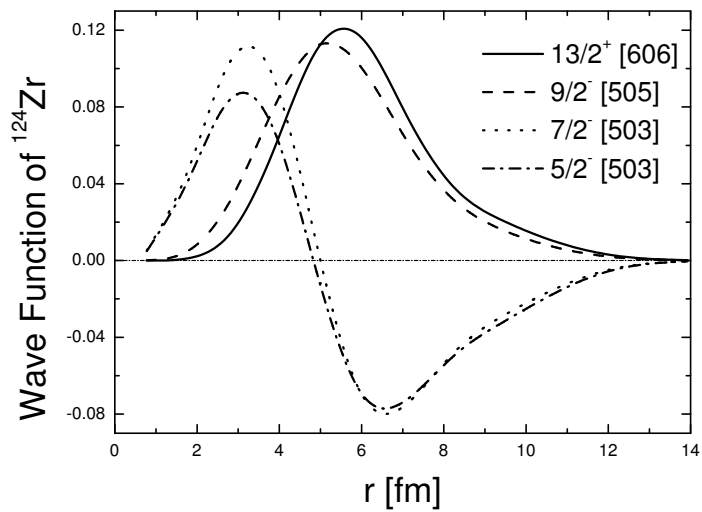}}
\caption{\label{fig5.fig} The wave function $f(z,r_\bot)$ for four
typical resonant states in $^{124}$Zr as functions of $r_\bot$ at
$z=0.422$. The corresponding energy (in MeV) and occupation
probability are, respectively, for $13/2^+ [606], E=5.276 \mbox{
and } v^2=0.005$; for $9/2^- [505], E=1.857\mbox{ and }
v^2=0.041$; for $7/2^- [503], E=0.464 \mbox{ and }v^2=0.196$ and
for $5/2^- [503], E=1.683 \mbox{ and }v^2=0.028$. The Nilsson
quantum numbers in square brackets are determined using the
dominant component in the expansion of this wave function in terms
of the anisotropic oscillator basis.}
\end{figure}
In Fig. \ref{fig4.fig}, we plot the occupation probabilities of
$^{124}$Zr for the neutron levels near the Fermi surface, i.e. in
the interval $-10\mbox{ MeV}\le E_{\textrm{s.p.}}\le 8$ MeV. We
present the occupation probabilities of neutron single-particle
states for two cases, the delta function interaction and the
constant pairing interaction usually used in the BCS framework. In
the constant pairing calculation, we used $G_n=12.0/A$ and
$G_p=30.0/A$, where $A$ is the mass number of $^{124}$Zr, and the
pairing window $\varepsilon_i-\lambda\le 2(41A^{-1/3})$
\cite{Gambhir90}. Here, we used different pairing strengths, $G_p$
and $G_n$, for protons and neutrons in order to obtain similar gap
energies, $\Delta$, than those obtained with the delta function
interaction. We note that we use different pairing strengths for
protons and neutrons because the level density around the Fermi
surface is small for protons and large for neutrons, due to the
continuum. The results obtained from the calculations using the
delta function interaction and the constant pairing interaction
are plotted in the left and right panels, respectively. Because in
our calculation $^{124}$Zr is almost spherical, we denote the
corresponding spherical quantum number in the left panel. We can
see all the spherical shell model states with small spin-orbit
splitting in the positive energy region.  They are all resonance
states supported by the centrifugal potential. A recently
performed resonant RMF+BCS calculation \cite{sand.03} yielded  a
quite similar spectrum for $^{124}$Zr. It is thus seen that the
contributions from the continuum states are very important. We see
clearly that the BCS calculation with the delta function
interaction can include more of the continuum states that are more
localized inside the nuclear region, which correspond to resonance
states. In addition, the delta function interaction includes a
smaller contribution of the continuum states that are less
localized in the pairing calculations. To see this point more
clearly, we plot the corresponding resonant wave function $f$ in
Fig. \ref{fig5.fig}. It is easily seen that these states do have
similar behavior as bound states. On the other hand, in the
constant pairing case, the occupation probabilities decrease
monotonically as functions of the single-particle energy.

We next point out an interesting feature of the vacancy between
$E\approx 2 \mbox{MeV}$ and $E\approx 5 \mbox{MeV}$ in the
single-particle spectra in the continuum seen in both
calculations. If we performed calculations for the single-particle
states in the coordinate space using box boundary conditions
instead of the harmonic oscillator expansion method, we would
obtain many states in the continuum.  These are the so-called
scattering states, which have small probabilities in the nuclear
region.  We do not obtain these scattering states, at least in the
region of several MeV excitation energy in the continuum. Hence,
we find that the single-particle states near the continuum
threshold do have large probabilities in the nuclear region and
contribute to the pairing correlations in nuclei close to the
neutron drip line.

This unique feature of the delta function interaction in the BCS
method is essential for the study of drip-line nuclei, where the
Fermi energy is close to the threshold of the continuum.  In this
case, we have to estimate correctly the coupling between the bound
states and the continuum states in order to pick up the resonance
states, which have large amplitudes in the nuclear region. This
would justify the use of such a simple RMF+BCS model to study all
the nuclei, including the unstable ones from the proton drip line
to the neutron drip line, as already demonstrated for the
spherical case by Yadav et al. \cite{yadav02}.

\subsection{Effect of pairing on the deformation}
The quadrupole moments of proton, neutron and nucleon
distributions are calculated as
\begin{equation}
\begin{array}{lll}
Q_i&=&\sqrt{\frac{\D16\pi}{\D5}}\langle r^2Y_{20}(\theta)\rangle_i\\
&=&\langle 2z^2-x^2-y^2\rangle_i,\\
\end{array}
\end{equation}
where $Y_{lm}$ represents the spherical harmonics, with $l$ being
the multi-polarity. The index $i=p,n,m$ denotes the expectation
value with respect to the proton, neutron and nucleon
distributions, respectively.

The deformation parameters are defined in terms of the liquid drop
model with uniform density. We expand the sharp surface of the
liquid drop as
\begin{equation}
R_i(\theta)=R_0\left[1+\beta_{2i}Y_{20}(\theta)+\beta_{4i}Y_{40}(\theta)\right],
\end{equation}
where $R_0=1.2A^{1/3}$ fm, in terms of the spherical harmonics
under the assumption of axial symmetry. The quadrupole moment of
the nucleon distribution in the liquid model is calculated as
\begin{equation}
Q_m^{\textrm{liq}}=\sqrt{\frac{\D16\pi}{\D5}}\frac{\D3A}{\D4\pi}R^2_0\beta_{2m},
\end{equation}
dropping terms of higher order in $\beta_{\lambda m}$. The
deformation parameter $\beta_{2m}$ is determined such that
$Q_m^{\textrm{liq}}$ reproduces $Q_m$ in the RMF calculation, and
is given by
\begin{equation}
\begin{array}{lll}
\beta_{2m}&=&\frac{\D\sqrt{5\pi}}{\D3}\frac{\D1}{\D AR^2_0}Q_m\\
&=&\frac{\D\sqrt{5\pi}}{\D3}\frac{\D1}{\D AR^2_0}\langle
2z^2-x^2-y^2\rangle_m.
\end{array}
\end{equation}
The deformation parameters $\beta_{2p}$ and $\beta_{2n}$ are
similarly given by \begin{equation}
\beta_{2p}=\frac{\D\sqrt{5\pi}}{3}\frac{\D1}{\D
ZR^2_0}Q_p,\end{equation}
\begin{equation}
\beta_{2n}=\frac{\D\sqrt{5\pi}}{3}\frac{\D1}{\D
NR^2_0}Q_n.\end{equation}

In the procedure to extract $\beta_{\lambda i}$ described above,
we retain the linear relations between the moments and the
deformation parameters so that it is easy to reproduce the moments
from the deformation parameters in Table. \ref{table1}. Regarding
the determination of the deformation parameters with terms of
higher order in $\beta_{\lambda i}$, we refer the reader to the
description in Ref.\cite{Tajima.96}.
\begin{figure}\epsfxsize = 10 cm \centerline{
\includegraphics[width=0.8\textwidth]{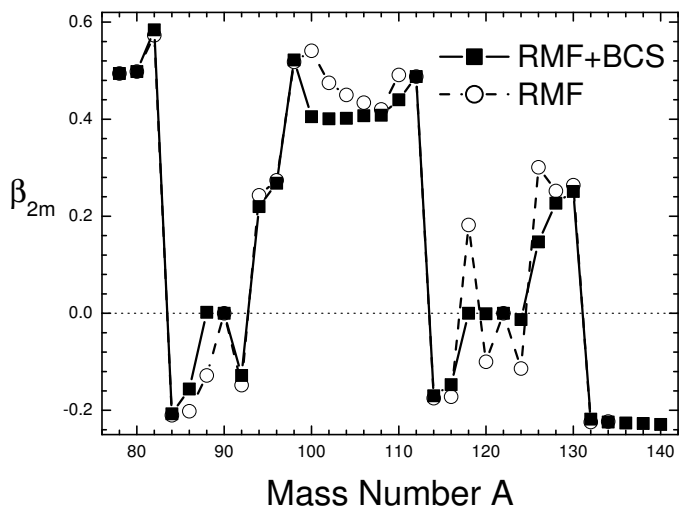}}
\caption{\label{fig6.fig} The quadrupole deformation,
$\beta_{2m}$, for even Zr isotopes obtained from the deformed
RMF+BCS (squares) and the deformed RMF (circles) calculations as
functions of mass number $A$.}
\end{figure}

\begin{figure}\epsfxsize = 10 cm \centerline{
\includegraphics[width=0.8\textwidth]{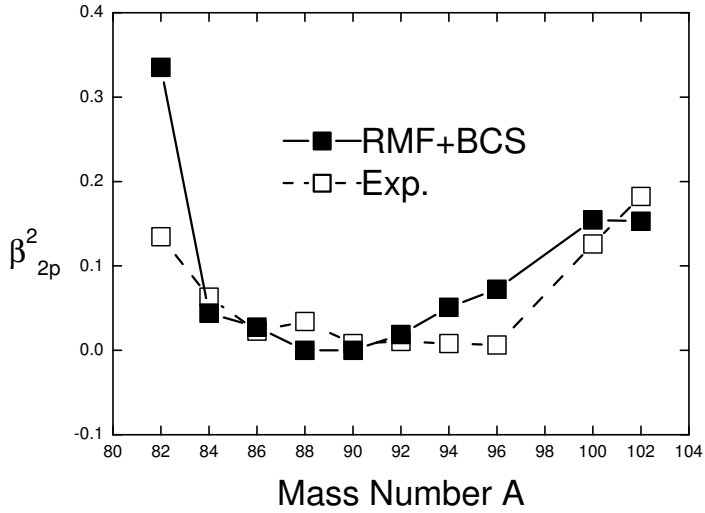}}
\caption{\label{fig7.fig} The square of the quadrupole
deformation,
 $\beta^2_{2p}$, for even Zr isotopes obtained from the deformed
RMF+BCS calculations (solid squares) and the empirical values
\cite{raman.01} (empty squares) as functions of mass number $A$.}
\end{figure}

 We plot in Fig. \ref{fig6.fig} the quadrupole deformation, $\beta_{2m}$,
 obtained from both the deformed RMF calculation and the deformed RMF+BCS
calculation. It is easily seen that the pairing effect reduces the
deformation of nuclei. More specifically, for some nuclei with
$|\beta_{2m}|< 0.2$ as found from the RMF calculation, (i.e.
$^{88}$Zr and $^{118}$Zr$-$$^{124}$Zr), the deformations are
removed almost completely by the pairing effect. For other largely
deformed nuclei (with $\beta_{2m}>0.3$), $\beta_{2m}$ is reduced
somewhat, but not as much as in the case of their weakly deformed
counterparts. In Fig. \ref{fig7.fig}, we compared our predictions
for the quadrupole deformation parameter $\beta^2_{2p}$ with the
empirical values \cite{raman.01}. Except for $^{82}$Zr, in which
case our result is larger than the empirical value, the results
are quantitively in good agreement with the empirical values.

\section{Conclusion}
We have formulated the RMF theory with deformation and pairing.
Conventionally, calculations for deformed nuclei are carried out
using the expansion method in terms of the harmonic oscillator
wave functions with pairing correlations treated using a constant
pairing interaction with a pairing window.  This method is,
however, not applicable to the case of nuclei close to the neutron
and proton drip lines, due to the importance of the resonance
states in the continuum for such nuclei. In order to extend the
RMF method so that it is applicable in these regions also, we
introduced a delta function interaction for the pairing
interaction, which allows the model to pick up resonant states by
making the pairing matrix elements state dependent. The delta
function description has been demonstrated to work for spherical
nuclei.

In this paper, to demonstrate the applicability of the method, we
have studied the Zr isotopes from the proton drip line to the
neutron drip line. We calculated the binding energies, nuclear
radii and deformation parameters of these nuclei. We also
calculated the binding energy per nucleon and the two-neutron
separation energy. We found that the agreement with experimental
values is very satisfactory. The effect of deformation is clearly
seen in these quantities as the large variation of the deformation
as a function of mass number.  We found that the neutron radii
increase monotonically with the neutron number, with some
anomalies, where the deformation changes suddenly. In our results,
the so-called giant halo effect is preserved for nuclei close to
the neutron drip line. The comparison of our results for
$\beta^2_{2p}$ with the empirical values shows that our
predictions for deformation parameters are quantitively in good
agreement with the experimental results.

The occupation probabilities in the continuum are important for
the purpose of determining if the present pairing method is
effective in the treatment of the pairing in the continuum. While
the occupation probabilities decrease monotonically as the states
deviate from the Fermi energy in the case of a constant pairing
interaction, they exhibit characteristic behavior for the case of
a delta function interaction.  The occupation probabilities are
large for those states whose wave functions have large overlap
with the wave functions below the Fermi surface.  It would be
interesting to compare our results with those of the
Hartree-Bogoliubov calculations for these nuclei.

The deformation varies greatly for the Zr isotopes.  The pairing
correlations have the effect of reducing the deformation. This is
clearly seen in our calculations.  For those nuclei that have
small deformations when calculated without pairing, the pairing
correlations cause the nuclei to be spherical.

In conclusion, we have carried out calculations for the Zr
isotopes to demonstrate that the presently considered RMF method
with deformation and pairing correlations is effective even for
nuclei close to the drip lines.  It would be very interesting to
make similar calculations for many nuclei in all mass regions with
the present method.

\section{Acknowledgements}
We acknowledge fruitful discussions and collaborations with Prof.
N. Sandulescu on the treatment of pairing in the continuum.

\end{document}